\documentclass[a4paper,twocolumn]{article}
\makeatletter
\def\thanks#1{\protected@xdef\@thanks{\@thanks
        \protect\footnotetext{#1}}}
\makeatother
\usepackage{amsmath,amssymb,amsfonts}
\usepackage{graphicx}
\usepackage{multirow}
\usepackage{threeparttable}
\usepackage[backend=biber,style=ieee,]{biblatex}
\usepackage{newtxtext,newtxmath}
\usepackage[center]{caption}
\usepackage{geometry}
\usepackage{authblk}

\title{End-to-end multi-channel speaker extraction and binaural speech synthesis}

\author[1,2]{Cheng Chi}
\author[1,3]{Xiaoyu Li}
\author[1,2]{Yuxuan Ke}
\author[4]{Qunping Ni}
\author[5]{Yao Ge}
\author[1,2]{Xiaodong Li}
\author[1,2,*]{Chengshi Zheng\thanks{*Corresponding author. E-mail: cszheng@mail.ioa.ac.cn}}

\affil[1]{Laboratory of Noise and Audio Research, Institute of Acoustics, Chinese Academy of Sciences, Beijing, China}
\affil[2]{University of Chinese Academy of Sciences, Beijing, China}
\affil[3]{Communication University of China, Beijing, China}
\affil[4]{Tianjin 712 Communication \& Broadcasting Co., Ltd.}
\affil[5]{Tianjin Tianan Borui Technology Co., Ltd.}

\date{}

\begin{document}

\twocolumn[
  \begin{@twocolumnfalse}
    \maketitle
    \begin{abstract}
      Speech clarity and spatial audio immersion are the two most critical factors in enhancing remote conferencing experiences. Existing methods are often limited: either due to the lack of spatial information when using only one microphone, or because their performance is highly dependent on the accuracy of direction-of-arrival estimation when using microphone array. To overcome this issue, we introduce an end-to-end deep learning framework that has the capacity of mapping multi-channel noisy and reverberant signals to clean and spatialized binaural speech directly. This framework unifies source extraction, noise suppression, and binaural rendering into one network. In this framework, a novel magnitude-weighted interaural level difference loss function is proposed that aims to improve the accuracy of spatial rendering. Extensive evaluations show that our method outperforms established baselines in terms of both speech quality and spatial fidelity.
    \end{abstract}
  \end{@twocolumnfalse}
]

\section{Introduction}
\label{sec:intro}
Binaural speech spatialization technique is crucial to improve the immersive experience of a remote conference for attendees when using VR/AR devices~\cite{gan2018natural, zhong2022binaural, jianjun2015natural}. Besides, because there often exist various noise sources in conference scenarios, such as air conditioning noise, typing noise, and whispering noise, speaker extraction is also highly demanded to improve speech clarity and intelligibility~\cite{hameed2021will,hawley2004benefit, middlebrooks2015sound}. This paper focuses on the tasks of speaker extraction and binaural speech synthesis to guarantee the quality of remote conferences. 

So far, many efforts have been made to achieve this goal. Among these efforts, single-channel-to-binaural synthesis has demonstrated impressive results~\cite{liu2022dopplerbas, liang2025binauralflow, parida2022beyond}. For this type of methods, their practical utility is often constrained by their reliance on auxiliary information, such as visual data or pre-defined source positions.~\cite{richard2020neural,gao20192,morgado2018self, kim2022immersive}.

Alternatively, microphone array offers a more self-contained solution because they can conduct speaker localization and speech separation without requiring auxiliary data. Using microphone array can capture spatial information from the soundfield  directly~\cite{he20243s,tian2023low}, which makes it popular for creating immersive experiences in telepresence applications. Traditionally, multi-channel-based systems obtain binaural speech through a multi-stage pipeline: at first, the direction-of-arrival (DOA) of the desired speech source is estimated~\cite{middlebrooks2015sound, krishnaveni2013beamforming}, and beamforming is then applied to reduce the background noise as well as the interference~\cite{Li2022TaylorBeamformerLA, 9746432}. Finally, the head-related transfer function (HRTF) is utilized to render the extracted speech into binaural signals~\cite{cobos2022overview,rafaely2022spatial,beit2022audio}. Nevertheless, this cascaded method has inherent limitations, including suboptimal noise reduction performance and the propagation of errors from early stages, such as inaccurate DOA estimation, which can degrade both the speech quality and the accuracy of spatial rendering.

\begin{figure}[t]
	\centering
	\includegraphics[width=1.0\columnwidth]{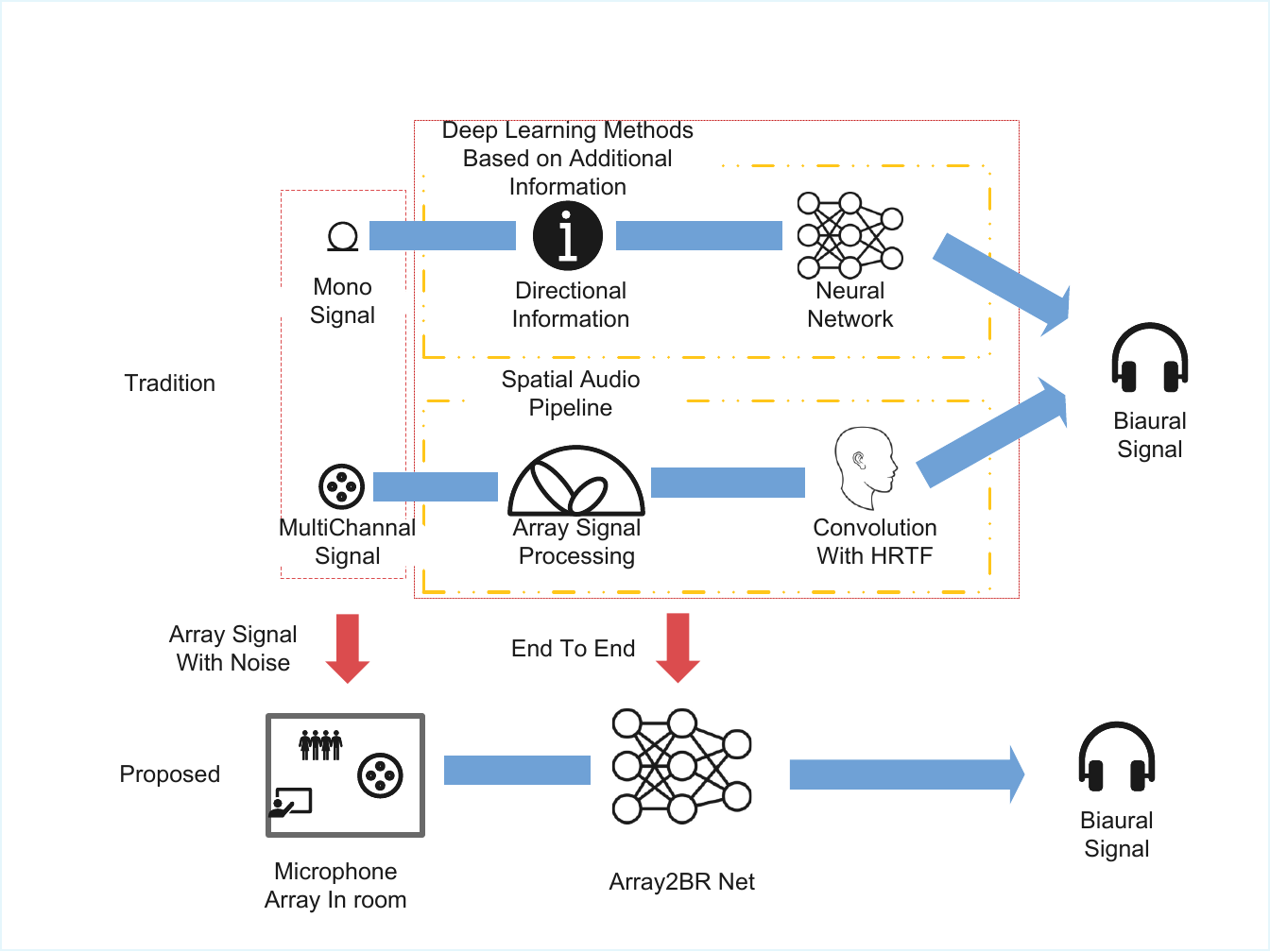}
	\caption{The diagram of the proposed framework.}
	\label{fig:framework}
\end{figure}

\begin{figure*}[t]
	\centering
	\setlength{\abovecaptionskip}{0.2cm}
	\setlength{\belowcaptionskip}{0.2cm}
	\centerline{\includegraphics[width=1.8\columnwidth]{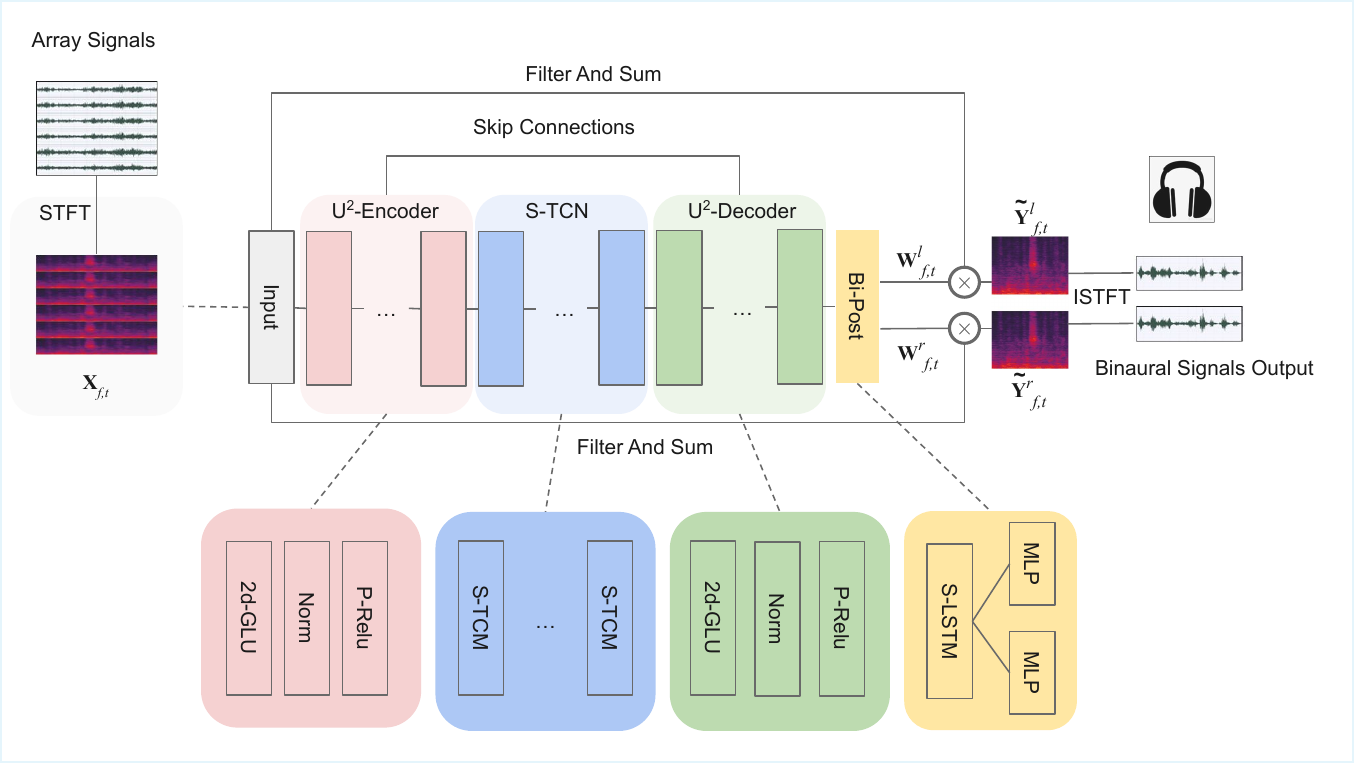}}
	\caption{The overall diagram of the proposed architecture.}
	\label{fig:architecture}
\end{figure*}

To overcome the above-mentioned limitations, we propose an end-to-end framework that directly synthesizes binaural speech from multi-channel noisy and reverberant signals. As illustrated in Fig.~\ref{fig:framework}, our method replaces the traditional multi-stage pipeline with a single network. This network learns the entire process as one unified task, simultaneously handling speaker localization, noise suppression, and binaural synthesis. The main contributions of this work are summarized as follows:
\begin{itemize}
    \item We propose a novel end-to-end architecture that directly synthesizes clean binaural speech from noisy multi-channel signals, unifying speaker extraction and spatial rendering.
    \item We introduce a magnitude-weighted Interaural Level Difference (mwILD) loss function, improving the preservation of spatial cues by focusing on perceptually significant components of the signal~\cite{nagel2018dynamic}.
    \item We conduct a comprehensive evaluation and show that our method outperforms established baselines in tests, achieving superior speech quality and spatial accuracy.
\end{itemize}

The remainder of the paper is organized as follows. Section~{\ref{sec:problem-formulation}} formulates the problem. Section~{\ref{sec:proposed-method}} elaborates on the proposed model, and its loss function is also presented in detail. In Sec.~{\ref{sec:experimental-setup}}, the configurations of the simulation experiment are introduced, and in Sec.~{\ref{sec:results-and-analysis}}, results and analysis are presented. In Sec.~{\ref{sec:conclusions}}, some conclusions are drawn.

\section{Problem Formulation}
\label{sec:problem-formulation}
In the context of remote meetings, which often involves a local and one or more remote sites, the quality of telepresence can be enhanced by audio processing. We consider a scenario where both a target speech source and ambient background noise are present. An \textit{M}-channel microphone array is employed to capture the sound field. The observed signal in the short-time Fourier transform (STFT) domain~\cite{allen1979image,stephenson1990comparison} can be modeled as:

\begin{equation}
  \label{frevec}
  \mathbf{X}_{f,t} = \mathbf{c}_fS_{f,t} + \mathbf{N}_{f,t},
\end{equation}
where $f \in \{1, \dots, F\}$ and $t \in \{1, \dots, T\}$ are the frequency and time indices, respectively. $\mathbf{X}_{f,t} \in \mathbb{C}^{M \times 1}$ is the complex-valued vector for the observed signal at the array. $\mathbf{c_f} \in \mathbb{C}^{M \times 1}$ represents the transfer functions from the speech source to the array microphones, and $S_{f,t}$ is the target speech signals. $\mathbf{N}_{f,t} \in \mathbb{C}^{M \times 1}$ is the complex-valued vector representing the ambient noise field captured by the array.

The desired output of our system, the clean binaural signals $\mathbf{Y}_{f,t}$, can be expressed as:

  \begin{equation}
    \begin{aligned}
        \mathbf{Y}_{f,t} &= \mathbf{a}_{f}{S}_{f,t}, \\
        \mathbf{Y}_{f,t} &= \left[Y^l_{f,t}, Y^r_{f,t}\right]^T,
    \end{aligned}
    \label{frevectar}
  \end{equation}

  \noindent where $\mathbf{a}_{f} \in \mathbb{C}^{2 \times 1}$ represents HRTFs, and the superscripts '\textit{l}' and '\textit{r}' in $Y^l_{f,t}$ and $Y^r_{f,t}$ denote the left and right ear signals. Our goal is to design a neural network that directly maps the noisy and reverberant  array observations $\mathbf{X}_{f,t}$ to the clean binaural signals $\mathbf{Y}_{f,t}$.

\section{Proposed Method}
\label{sec:proposed-method}
\subsection{Network Architecture}
Our proposed network, illustrated in Fig.~\ref{fig:architecture}, is an end-to-end U-Net-based architecture~\cite{qin2020u2} that is designed to directly map noisy multi-channel signals to clean binaural speech. The core of our model is a symmetric encoder-decoder structure enhanced with a specialized bottleneck for spatio-temporal modeling and a binaural synthesis head.

The process begins by transforming the \textit{M}-channel input signals into the time-frequency (T-F) domain using STFT. The complex spectra of the multi-channel input signals are fed into the U$^2$-Encoder. The encoder consists of a series of downsampling blocks, and each block employs a 2D Gated Linear Unit (2D-GLU) to capture significant local patterns in the spectro-temporal domain, followed by normalization and P-ReLU activation~\cite{he2015delving} for feature extraction.

At the bottleneck of the U-Net, we introduce spatial-temporal correlation network (S-TCN). This module is composed of several spatial-temporal correlation module (S-TCM) blocks, specifically designed to model the intricate dependencies across microphone channels and time frames~\cite{bai2018empirical}. By effectively processing these correlations, the S-TCN can disentangle the spatial and temporal characteristics of the desired speaker from background noise and interference.

The U$^2$-Decoder then reconstructs the signal representation. It mirrors the structure of the encoder, progressively upsampling the features while integrating high-resolution information from the encoder via skip connections. This U-Net topology is crucial for preserving the details necessary for speech synthesis.

Finally, instead of directly outputting spectrograms, the output of the encoder is passed to a binaural post-processing (Bi-Post) module. This module features two parallel branches, and each branch comprises a spatial LSTM (S-LSTM) and a multi-layer perceptron (MLP). These two branches generate two distinct complex-valued filters, $\mathbf{W}^{l}_{f,t} \in \mathbb{C}^{M \times 1}$ and $\mathbf{W}^{r}_{f,t} \in \mathbb{C}^{M \times 1}$, for each time-frequency (T-F) bin. The final binaural signals are synthesized by applying these filters to the corresponding T-F vectors of the multi-channel input, $\mathbf{X}_{f,t}$. The synthesis process can be expressed as:
  \begin{equation}
    \begin{aligned}
      \tilde{Y}^{l|r}_{f,t} =( \mathbf{W}^{l|r}_{f,t})^H \mathbf{X}_{f,t}
    \end{aligned}
  \end{equation}
  \noindent where $\tilde{Y}^{l|r}_{f,t}$ is the estimated binaural speech signal, with $l,r$ denoting the left or the right ear. The final two-channel time-domain signal is then obtained via Inverse STFT (ISTFT). Note that the entire architecture is trained in an end-to-end manner, allowing the network to jointly optimize for speaker extraction and accurate spatial rendering.

\subsection{Loss function}
\label{Loss function}
To ensure both high speech quality and accurate spatial rendering, we design a composite loss function, which is a weighted sum of three different components: a real-imaginary (RI) loss, a magnitude (Mag) loss, and the proposed magnitude-weighted interaural level difference (mwILD) loss. Thus, the loss function is defined as:
\begin{equation}
  \label{eq:loss_final}
  \mathcal L = \lambda_1\mathcal L_{RI} + \lambda_2 \mathcal L_{Mag} + \lambda_3 \mathcal L_{mwILD}.
\end{equation}
  
In ~(\ref{eq:loss_final}), the $\mathcal{L}_{RI}$ and $\mathcal{L}_{Mag}$ are defined as follows:
\begin{equation}
  \begin{aligned}
    \mathcal{L}_{RI}(\tilde{\mathbf{Y}}, \mathbf{Y}) &= \|\tilde{\mathbf{Y}}_r-\mathbf{Y}_r\|_F^2+\|\tilde{\mathbf{Y}}_i-\mathbf{Y}_i\|_F^2, \\
    \mathcal{L}_{Mag}(\tilde{\mathbf{Y}}, \mathbf{Y}) &= \|\sqrt{|\tilde{\mathbf{Y}}_r|^2+|\tilde{\mathbf{Y}}_i|^2}-\sqrt{|\mathbf{Y}_r|^2+|\mathbf{Y}_i|^2}\|_F^2,
  \end{aligned}
\end{equation}
\noindent where the time and frequency indices $(f,t)$ are omitted for brevity when no confusion arises. $\tilde{\mathbf{Y}} = [\tilde{Y}^l, \tilde{Y}^r]^T$ and $\mathbf{Y} = [Y^l, Y^r]^T$ represent the estimated and target binaural complex spectra, respectively. $\{\tilde{\mathbf{Y}}_r, \tilde{\mathbf{Y}}_i\}$ and $\{\mathbf{Y}_r, \mathbf{Y}_i\}$ denote the real and imaginary parts of $\tilde{\mathbf{Y}}$ and $\mathbf{Y}$, respectively. $||\cdot||_F$ denotes the Frobenius norm.
The mwILD loss is defined as:
\begin{equation}
  \mathcal L_{mwILD}(\tilde{\mathbf{Y}}, \mathbf{Y}) =  \left\| \frac{\sum\limits_{f,t}\sigma_{f,t} \left( \mathrm{ILD} (\tilde{\mathbf{Y}}_{f,t})-\mathrm{ILD} (\mathbf{Y}_{f,t}) \right) }{\sum\limits_{f,t}\sigma_{f,t}} \right\|_1.
\end{equation}
\noindent Note that a conventional ILD loss treats all time-frequency (T-F) bins uniformly, and this loss can be inaccurate. For example, a near-zero overall ILD might mask significantly but opposing ILD variations across different frequency bands, providing an inadequate learning signal for the network. To align the loss with this physical principle, the proposed mwILD loss introduces an energy-based weighting mechanism. Here, the weight $\sigma_{f,t}$ is the sum of the energy from the left and right channels of the desired signal. This approach scales the ILD error at each T-F bin, prioritizing spatial accuracy in the most perceptually salient, high-energy components. The L1 norm is applied to the final error term to enhance robustness against outliers. The weights $\left \{ \lambda_{1}, \lambda_{2}, \lambda_{3} \right \}$ in (\ref{eq:loss_final}) are empirically set to $\left \{ 1, 1, 3 \right \}$ to balance the contributions of each term.
\begin{figure}[t]
	\centering
  \vspace{-0.2cm} 
	\setlength{\abovecaptionskip}{0.2cm}
	\setlength{\belowcaptionskip}{0.5cm}
	\centerline{\includegraphics[width=0.8\columnwidth]{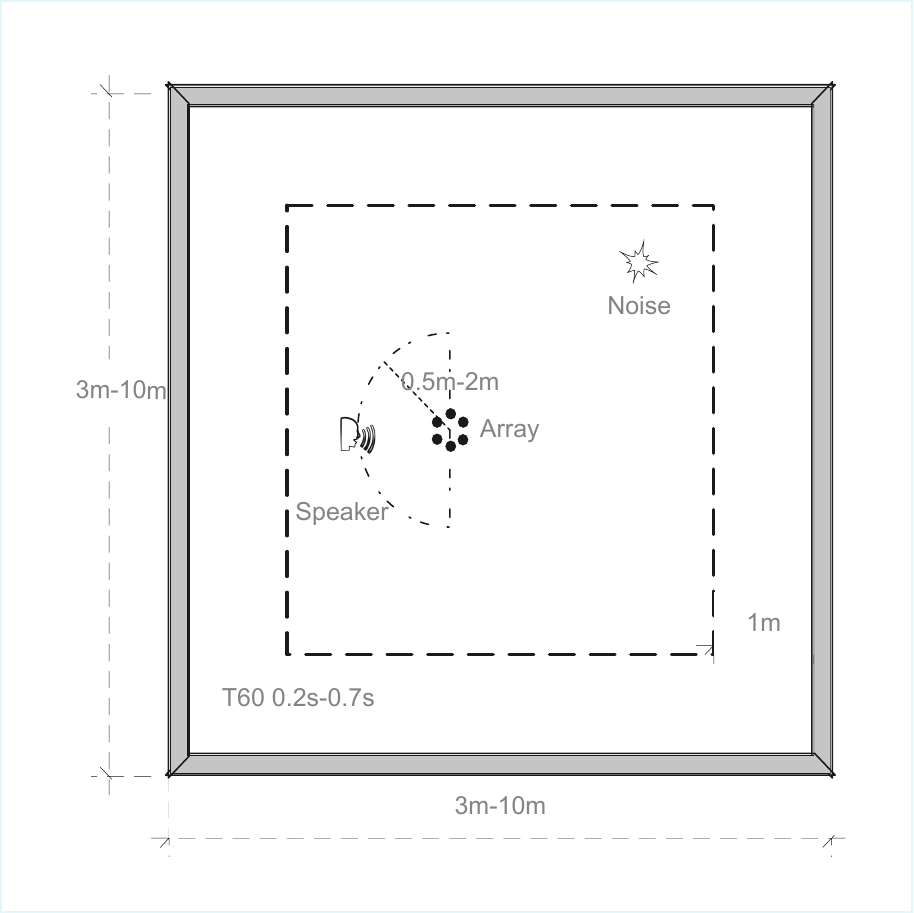}}
	\caption{Spatial configurations of the room, showing the random placement of sound sources within the designated area and the minimum distance requirements from walls.}
	\label{fig:room_settings}
	\vspace{-0.2cm}
\end{figure}

\begin{table*}[ht]
  \caption{Objective evaluation of different models on the test set, averaged across all SNR conditions. Our proposed model demonstrates superior performance with significantly lower complexity.}
  \vspace{0.2 cm}
  \small
  \centering
  \renewcommand\arraystretch{1.2} 
  \begin{tabular}{c | c | c | c c c c c}
      \hline
      \hline
      \textbf{Model} & \textbf{Params. [M]}$\downarrow$ & \textbf{FLOPs [G/s]}$\downarrow$ & $\mathbf{\Delta ITD}$ [ms]$\downarrow$  & $\mathbf{\Delta ILD}$ [dB]$\downarrow$  & \textbf{PESQ}$\uparrow$ & \textbf{ESTOI}$\uparrow$ & \textbf{SD [dB]}$\downarrow$ \\
      \hline
      LBH-MVDR  & - & - & - & - & 2.31 & 0.54 & 0.72 \\
      LBH-NN    & 2.16 & 0.89 & - & - & 2.98 & 0.79 & 0.15 \\
      MIF       & - & - & 0.250 & 1.06 & 1.96 & 0.45 & 40.25 \\
      MDFnet    & 6.36 & 2.14 & 0.012 & 0.74 & 2.41 & 0.55 & 0.44 \\
      Proposed & 2.37 & 1.01 & 0.002 & 0.21 & 2.89 & 0.71 & 0.23 \\
      \hline
      \hline
  \end{tabular}
  \label{tab:params_flops}
\end{table*}

\section{Experimental Setup}
\label{sec:experimental-setup}
\subsection{Dataset configuration}
For comprehensive evaluation of the proposed method, we constructed a large-scale dataset using the DNS-Challenge corpus~\cite{reddy2020interspeech}. The dataset was partitioned into training, validation, and testing sets with the ratio of 20:2:1. A six-element uniform circular array (UCA) with a diameter of 8 cm was employed to capture the spatial acoustic information~\cite{aboumahmoud2021review}.

To create acoustically diverse scenarios, we simulated various room configurations and obtained the simulated room impulse responses (RIRs) by using the image method~\cite{allen1979image,habets2006room,lehmann2008prediction}. Fig.~\ref{fig:room_settings} illustrates the spatial setup of our simulated environment. The room dimensions were randomly sampled between 3~m and 10~m for both length and width, while the reverberation time was set from 0.2~s to 0.7~s to represent different acoustic environments. Considering the boundary effects, both the sound sources and the microphone array were placed at least 1 m away from the wall, and the area of them was indicated by a dashed line in Fig.~\ref{fig:room_settings}. Besides, the distance between the target sound source and the array was set randomlly from 0.5 m to 2 m.

To simulate realistic conference scenarios, the noise was selected from the DNS-Challenge noise corpus. Their distances from the array ranged from 0.5 m to 2 m, and their azimuth range was set from $-90^{\circ}$ to $90^{\circ}$. The signal-to-noise ratio (SNR) varied from 0~dB to 30~dB to evaluate system performance under different noise conditions. For each configuration, we generated the mixed signals by convolving clean speech with the simulated RIRs, while the target binaural signals were synthesized by applying HRTFs~\cite{evans1998analyzing} corresponding to the source directions.

\subsection{Training configuration}
In our experiment, all the utterances were sampled at 16~kHz. For STFT, a squared-root Hanning window was selected between adjacent frames with 50~\% overlap, so the frame length and frame shift were, respectively, set to 20 ms and 10 ms. Adam optimizer was utilized to train the model~\cite{kingma2014adam}. The initial learning rate was set to 5e-4, which would be halved if the loss value does not decrease for three continuous epochs. The training procedure would stop once the learning rate halves for fourth time. Both the training and the testing of neural networks were conducted using a Nvidia Tesla V100 GPU.

\section{Results and Analysis}
\label{sec:results-and-analysis}
We compared our proposed framework against both conventional and deep learning-based methods. The conventional baselines include: 1) LBH-MVDR~\cite{beit2022audio}, a classic pipeline involving Localization, Beamforming, and HRTF filtering, with MVDR used as beamforming; 2) LBH-NN, which used the same architecture as our proposed model but with only a single MLP branch for beamforming to generate monaural audio~\cite{zheng2023sixty}, followed by HRTF filtering; and 3) MIF~\cite{hsu2023model}, a multi-channel inverse filtering method. As for the deep learning-based baseline, 4) MDFNet~\cite{miyoshi1988inverse}, a recent high-quality end-to-end model, was also chosen.

We employed seven objective metrics to conduct a comprehensive evaluation. Model complexity was measured by the number of Parameters and FLOPs. Spatialization accuracy was assessed by the absolute error in interaural time difference ($\Delta$ITD) and interaural level difference ($\Delta$ILD)~\cite{eason1955certain}. Speech quality and intelligibility were evaluated using perceptual evaluation of speech quality (PESQ)~\cite{geiser2007bandwidth} and extended short-time objective intelligibility (ESTOI)~\cite{jensen2016algorithm}. Finally, the spectral distortion between the synthesized and target signals was measured by spectral distance (SD). The evaluation was performed on a test set with new, randomly generated room configurations, and the results were averaged across all SNR conditions.

As shown in Table~\ref{tab:params_flops}, our proposed model demonstrates a superior balance between performance and computational efficiency. With 2.37~M parameters and 1.01~G/s FLOPs, it is more lightweight than MDFnet. In terms of spatial accuracy, our model surpassed all baselines, achieving the $\Delta$ITD (0.002~ms) and $\Delta$ILD (0.21~dB). For perceptual quality and intelligibility, it scored a competitive PESQ of 2.89 and an ESTOI of 0.71. Furthermore, it obtained a low SD of 0.23~dB, indicating high fidelity in spectral reconstruction. While the LBH-NN pipeline achieved the best scores on speech quality metrics, it requires external DOA information and is not an end-to-end system. In contrast, our method achieved comparable performance without such auxiliary information, highlighting its practical advantages. MIF had a much higher spectral distance, and this is because this method does not include a denoising component. In summary, our framework establishes a new state-of-the-art by delivering superior spatialization with a highly efficient model, crucially without the need for external directional information.

Regarding spatial rendering accuracy, the results of the LBH-based methods are not presented in Table~\ref{tab:params_flops} as their spatialization is the ground-truth. These methods rely on oracle directional information for HRTF filtering, thus introducing no spatial rendering error. Among the end-to-end models that learn to spatialize, our proposed method demonstrates superior performance. As shown in Table~\ref{tab:params_flops}, our method achieves much lower $\Delta$ITD and $\Delta$ILD compared with both MIF and MDFnet. This highlights the effectiveness of our framework in learning and generating spatial cues directly from the multi-channel signals, without estimating the direction-of-arrival of the desired speaker explicitly.



\section{Conclusions}
\label{sec:conclusions}
In this paper, a novel end-to-end framework is proposed for generating clean and spatialized binaural speech directly from the multi-channel noisy and reverberant signals, which is achieved by unifying source extraction, noise suppression, and binaural rendering into a single network. A magnitude-weighted ILD loss function is also introduced to effectively enhance spatial rendering accuracy. Experimental results demonstrated that our method effectively improves both speech quality and spatial fidelity when compared with existing methods.

Despite its promising performance, our work has  many limitations. The current framework is designed to handle only one desired speaker. Real-world conferencing scenarios, however, often involve multiple simultaneous desired talkers. A key direction for future research is to extend our model to multi-talker scenarios. Addressing this challenge will be a crucial step toward creating truly immersive and practical telepresence systems.

\section{Acknowledgement}
The authors would like to express their sincere gratitude to Yicheng Hsu at National Tsing Hua University for his invaluable assistance and support in testing and evaluating the baseline models.

\printbibliography[heading=bibintoc]

\end{document}